\documentclass[12pt]{iopart}

\usepackage{iopams}
\usepackage{epsfig}
\usepackage{amssymb}

\def\ket#1{|\,#1\,\rangle}

\newcommand{\beq}{\begin{equation}}
\newcommand{\eeq}{\end{equation}}
\newcommand{\beqa}{\begin{eqnarray}}
\newcommand{\eeqa}{\end{eqnarray}}

\begin{document}

\title[Quantum Chernoff bound as a measure of efficiency of cloning]{Quantum Chernoff bound as a measure of efficiency of quantum cloning for mixed states}

\author{Iulia Ghiu}

\address{Centre for Advanced Quantum Physics,
Department of Physics, University of Bucharest,
P. O.  Box MG-11, R-077125  Bucharest-M\u{a}gurele, Romania}

\ead{iulia.ghiu@g.unibuc.ro}

\begin{abstract}
In this paper we investigate the efficiency of quantum cloning of $N$ identical mixed qubits. We employ a recently introduced measure of distinguishability of quantum states called quantum Chernoff bound. We evaluate the quantum Chernoff bound between the output clones generated by the cloning machine and the initial mixed qubit state. Our analysis is illustrated by performing numerical calculation of the quantum Chernoff bound for different scenarios that involves the number of initial qubits $N$ and the number of output imperfect copies $M$.
\end{abstract}


\maketitle

\section{Introduction}

The {\it no-cloning} theorem describes a protocol that proves one of the major differences between the classical information theory and the quantum information processing, where the use of quantum states and quantum operations are allowed. This theorem was given by Wootters and \.{Z}urek, who have shown that no linear operator exist that can generate two perfect copies of an unknown quantum state \cite{Wootters}.

Because the perfect cloning cannot be realized, there are many proposals of quantum cloning machines that produce imperfect copies. Such a machine is denoted by $N\to M$, which means that there are $N$ identical copies of the initial state and $M$ imperfect clones that are generated (with $M\ge N$).
Firstly, the analysis was focus on machines that are applied on pure states: $1\to 2$ cloning of qubits \cite{Bruss-1998}, $1\to 2$ cloning of $d$-dimensional systems \cite{Buzek-1998}, $N\to M$ cloning machine of qubits \cite{Gisin,Bruss-2} and qudits \cite{Werner,Zanardi, Fan-2001}. A further investigation was done for asymmetric cloning, this being a machine that generates two different clones whose Bloch vectors are parallel to the Bloch vector of the initial state to be cloned. Optimal universal asymmetric cloning machines were proposed by Cerf for qubits \cite{Cerf0}, as well as for $d$-dimensional systems \cite{Cerf1}.
Both symmetric and asymmetric cloning machine of pure states are of great importance in a process called quantum telecloning that enables the transmission of the imperfect clones at a distance \cite{Murao-1999}. This protocol was extended to the symmetric case for $d$-level systems and asymmetric telecloning of qubits \cite{Murao-2000, Ghiu-2005}, as well as asymmetric telecloning of qudits \cite{Ghiu-2003}. A different generalization of telecloning consists of using a non-maximally entangled channel as a resource \cite{Rigolin,Ghiu-2012,Ghiu-2013}.

Secondly, the optimal universal symmetric cloning of mixed states was investigated: $2\to M$ cloning \cite{Fan-q-inf} and $N\to M$ cloning \cite{Fan-pra}. In this paper we analyze the quality of the imperfect clones generated by cloning machines applied on mixed qubits. Our proposal is based on a recently introduced measure of distinguishability of quantum states called quantum Chernoff bound.

In 1952, Chernoff has proved a bound on the minimal error probability of discriminating two probability distributions in the asymptotic case \cite{Chernoff}. One of the properties of the classical Chernoff bound is that it is a natural distance between probability distributions. This bound has found many applications in statistical decision theory. After more than fifty years, the generalization to the quantum regime was proved by two groups of researchers: Nussbaum and Szko\l a, and Audenaert {\it et al.} \cite{Nussbaum,Acin}. In the case of quantum systems we start with $k$ identical copies of a quantum system, which is prepared in the same unknown state described by the density operators $\rho $ or $\zeta $. Further one has to determine the minimal probability of error by testing the copies in order to discover the identity of the state. The quantum Chernoff bound was employed in different branches of physics: quantum information theory, quantum optics, statistical physics, just to enumerate a few. It was used as a measure of distinguishability between mixed qubit states and between single-mode Gaussian states of the radiation field \cite{Calsa}. The quantum degree of polarization of a two-mode state of the quantum radiation filed was given in terms of the quantum Chernoff bound \cite{Ghiu-pra-2010,Ghiu-oc-2010}. Discrimination between two ground states or two thermal states of the one-dimensional quantum Ising model was recently addressed with the help of the quantum Chernoff bound by Invernizzi and Paris \cite{Paris}.

The paper is organized as follows. We start by reviewing the $N\to M$ cloning machine applied on mixed qubits in Sec. 2. Further in Sec. 3 we present the quantum Chernoff bound for mixed qubits. We evaluate the quantum Chernoff bound between the output clones generated by the cloning machine of Sec. 2 and the initial mixed qubit to be cloned. Our conclusions are outlined in Sec. 4.

\section{Cloning of mixed states}

\subsection{$2\to M$ cloning of qubits}

Suppose that we have two identical mixed qubits described by $\rho \otimes \rho $. Optimal universal cloning of this system was  investigated in Ref. \cite{Fan-q-inf}. If the output copies are independent on the initial state, then the cloning machine is called universal. In other words, the density operator of the clones must have the following expression
\beq
\rho_{out}=\eta \rho +\frac{1-\eta }{2}I,
\eeq
where $\eta $ is the shrinking factor which is independent of the input state $\rho $, and $I$ is the identity
operator. The cloning machine is optimal if the shrinking factor is maximal.

In Ref. \cite{Fan-q-inf}, Fan {\it et al.} have shown that the imperfect clones of the $2\to M$ optimal universal cloning machine are characterized by the density operator $\rho_{out}$:
\beq
\rho_{out}=\frac{M+2}{2M}\, \rho +\frac{M-2}{4M}\, I,
\eeq
$M$ being the number of the generated copies that satisfies $M\ge 2$.

\subsection{$N\to M$ optimal universal cloning of qubits}

The more general scenario is the one that considers $N$ identical mixed qubits $\rho^{\otimes N}$ as being the input state. The task is to generate $M$ imperfect clones, $M$ satisfying $M\ge N$. The proposal of $N\to M$ universal quantum broadcasting of mixed states was analyzed in Ref. \cite{Fan-pra}. Dang {\it et al.} have used a special decomposition of the $N$ identical mixed states by taking into account both the symmetric and asymmetric states.

Let us consider the following notations for the states with $(N-m)$ spins {\it up} and $m$ spins {\it down} \cite{Fan-pra}: $\left\vert \left( N-m\right) \uparrow
,m\downarrow \right\rangle _{\alpha =0}:= \left\vert \left( N-m\right) \uparrow
,m\downarrow \right\rangle $ for the symmetric state, that is invariant under the permutation of the particles. Further we denote the state $\left\vert \left( N-m\right) \uparrow ,m\downarrow \right\rangle
_{\alpha \ne 0}$ to be an asymmetric one, since it contains different phases. The general expression is given by:
\begin{eqnarray}
&&\left\vert \left( N-m\right) \uparrow ,m\downarrow \right\rangle _{\alpha }
\nonumber \\
&:= &\frac{1}{\sqrt{C_{N}^{m}}}\sum_{j=1}^{C_{N}^{m}}
e^{2\pi i\alpha \left( j-1\right) /C_{N}^{m}}\Pi _{j}\left( \left\vert
\uparrow \right\rangle ^{\otimes N-m}\left\vert \downarrow \right\rangle
^{\otimes m}\right) .
\label{stari}
\end{eqnarray}
$\Pi _{j}$ is the $j$-th permutation operator, while the number of the permutation operators is $C_{N}^{m}$.

As an example we take $N=3$ and $m=1$. According to Eq. (\ref{stari}), we obtain \cite{Fan-pra}:
\beqa
\left\vert
2\uparrow ,\downarrow \right\rangle _{0}&:=& \left\vert 2\uparrow
,\downarrow \right\rangle =\frac{1}{\sqrt{3}}\left( \left\vert \uparrow
\uparrow \downarrow \right\rangle +\left\vert \uparrow \downarrow \uparrow
\right\rangle +\left\vert \downarrow \uparrow \uparrow \right\rangle \right)\nonumber \\
\left\vert 2\uparrow ,\downarrow \right\rangle _{1}&:=&\frac{1}{\sqrt{3}}
\left( \left\vert \uparrow \uparrow \downarrow \right\rangle +\omega
\left\vert \uparrow \downarrow \uparrow \right\rangle +\omega ^{2}\left\vert
\downarrow \uparrow \uparrow \right\rangle \right)\nonumber \\
\left\vert 2\uparrow
,\downarrow \right\rangle _{2}&:=&\frac{1}{\sqrt{3}}\left( \left\vert \uparrow
\uparrow \downarrow \right\rangle +\omega ^{2}\left\vert \uparrow \downarrow
\uparrow \right\rangle +\omega \left\vert \downarrow \uparrow \uparrow
\right\rangle \right),
\eeqa
with $\omega =e^{2\pi i/3}$.

The spectral decomposition of an arbitrary mixed qubit state is given by
\begin{equation}
\rho =a\left\vert \uparrow \right\rangle \left\langle \uparrow
\right\vert +b\left\vert \downarrow \right\rangle \left\langle
\downarrow \right\vert ,
\end{equation}
where $a+b=1$.

By using the notations (\ref{stari}), Dang {\it et al.} have determined the general decomposition of
the state of $N$ identical mixed qubits \cite{Fan-pra}
\begin{eqnarray}
\rho ^{\otimes N} =\sum_{m=0}^N a^{N-m}\; b^{m}\sum_{\alpha =0}^{C_{N}^{m}-1} \left\vert
\left( N-m\right) \uparrow ,m\downarrow \right\rangle _{\alpha
\alpha }\left\langle \left( N-m\right) \uparrow ,m\downarrow
\right\vert .
\end{eqnarray}

The $N\to M$ optimal universal cloning machine was found in Ref. \cite{Fan-pra} and is defined by the unitary operator:
\begin{eqnarray}
&&U \left\vert \left( N-m\right) \uparrow ,m\downarrow
\right\rangle _{\alpha }\otimes \ket{R}  \nonumber \\
&=&\sum_{k=0}^{M-N} \beta _{mk}\left\vert
\left( M-m-k\right) \uparrow ,\left( m+k\right) \downarrow
\right\rangle _{\alpha } \otimes \ket{R_{\left\vert \left(
M-N-k\right) \uparrow ,k\downarrow \right\rangle _{\alpha }}}.
\label{tr-unitara}
\end{eqnarray}
In Eq. (\ref{tr-unitara}), $\ket{R}$ represents the input blank state of the ancilla, while $\ket{R_{\left\vert \left(M-N-k\right) \uparrow ,k\downarrow \right\rangle _{\alpha }}}$ is the state of the ancilla after the unitary operator was applied. The coefficients $\beta _{mk}$ are defined as
\begin{eqnarray}
\beta _{mk} :=\sqrt{\frac{\left( M-N\right) !\left( N+1\right)
!}{\left( M+1\right) !}}\sqrt{\frac{\left( M-m-k\right) !}{\left(
N-m\right) !\left( M-N-k\right) !}} \sqrt{\frac{\left( m+k\right)
!}{m!k!}}.
\end{eqnarray}

This cloning machine generates $M$ clones described by the following density operator \cite{Fan-pra}:
\beq
\rho_{out}=\frac{N\left( M+2\right) }{M\left( N+2\right) }\; \rho
+\frac{M-N}{M\left( N+2\right) }\; I.
\label{stare-clonata}
\eeq

In the next section, we investigate the efficiency of this cloning machine by using the quantum Chernoff bound. There are two different approaches: firstly $N$ is kept constant and we perform the analysis by varying $M$. Secondly, we take $M$ to be constant and evaluate what happens when $N$ has different values.

\section{Quantum Chernoff bound as a measure of efficiency of cloning. Numerical results}

\subsection{Quantum Chernoff bound for qubits}

Quantum Chernoff bound is a recently introduced measure that enables the discrimination of two quantum states, whose density operators are $\rho $ and $\zeta $. Suppose that $k$ identical copies of either $\rho $ or $\zeta $ are given to an observer. The task of the observer is to determine the minimal probability of error for identifying the quantum state
by performing quantum operations on the $k$ copies \cite{Acin}. In the particular case of equiprobable states, the minimal error probability of discriminating them in a measurement performed
on $k$ independent copies is \cite{Kargin,Auden}
\begin{equation}
P_{\rm min}^{(k)}(\rho,\, \zeta)=\frac{1}{2}\left(1-\frac{1}{2}
||{\rho}^{\otimes k}-{\zeta}^{\otimes k}||_1\right)
\label{error},
\end{equation}
where $||A||_1:=\mbox{Tr}\sqrt{A^{\dagger} A}$ is
the trace norm of a trace-class operator $A$. The right-hand side of Eq. (\ref{error}) is in general difficult to be evaluated. If the two states are both pure, let us denote them by $|\Phi \rangle$ and $|\Psi \rangle$, then the minimal probability (\ref{error}) has an analytical expression that involves only the transition probability between the two states \cite{Kargin}
\beq
P_{\rm min}^{(k)}(|\Phi \rangle \langle\Phi|, \;|\Psi \rangle \langle\Psi|)
=\frac{1}{2}\left( 1-\sqrt{1-|\langle \Phi \ket{\Psi}|^{2k}}\right).
\eeq

For a large number of identical copies, in the asymptotic limit, an upper bound
of the minimal probability of error (\ref{error}) was found to decrease exponentially with $k$ \cite{Acin}:
$$P^{(k)}_{\rm min}(\rho,\, \zeta)\sim \exp \left[-k\; \xi _{QCB}(\rho, \,\zeta)\right]. $$

The positive quantity  $\xi_{QCB}(\rho,\, \zeta)$ is called quantum Chernoff bound \cite{Nussbaum,Acin} and is defined as follows
\beqa
\xi_{QCB}(\rho,\, \zeta)&:=& \lim_{k\to \infty }-\frac{\ln P^{(k)}_{\rm min}(\rho,\, \zeta)}{k} \nonumber \\
&=&-\ln \left[ \min_{s\in [0,1]}\mbox{Tr}\left({\rho}^s{\zeta}^{1-s}\right) \right].
\label{xi}
\eeqa

The non-logarithmic variety of the quantum Chernoff bound is denoted in Ref. \cite{Acin} as
\begin{equation}
Q(\rho,\, \zeta):=\min_{s\in [0,1]}\mbox{Tr}({\rho}^s{\zeta}^{1-s}).
\label{QCB}
\end{equation}
This function is symmetric, i.e. $Q(\rho,\, \zeta)=Q(\zeta ,\, \rho),$ and is called the quantum Chernoff overlap of the states $\rho $ and $\zeta $  \cite{Ghiu-pra-2010}.

A different measure of distinguishability of two quantum states is the fidelity, that is defined as \cite{Jozsa}
\[
F(\rho,\, \zeta):=\bigg[ \mbox{Tr} \bigg( \sqrt{\sqrt{\rho }\; \zeta \, \sqrt{\rho }}\bigg) \bigg] ^2.
\]

There is an interesting connection between the quantum Chernoff bound, or more specifically the non-logarithmic variety of Chernoff bound $Q(\rho,\, \zeta)$, and the fidelity, namely the following two bounds are satisfied \cite{Calsa,Kargin}:
\[
F(\rho,\, \zeta)\le Q(\rho,\, \zeta) \le \sqrt{F(\rho,\, \zeta)}.
\]

Suppose that the two quantum states represent qubit mixed states. In other words, the two density operators are expressed in terms of the Bloch vectors $\vec r$ and $\vec p\; $ as
\beqa
\rho &=&\frac{1}{2}\bigg( I+\vec r\cdot \vec \sigma \bigg) \nonumber \\
\zeta &=&\frac{1}{2}\bigg( I+\vec p\cdot \vec \sigma \bigg),
\eeqa
$\vec \sigma $ being the Pauli operators and $r$, $p\in [0,1]$. The eigenvalues of the two density operators are the following:
\beqa
\mbox{for} \; \rho:&& \vspace{0.4cm} \lambda =\frac{1}{2}(1+r); \; \; \tilde \lambda =\frac{1}{2}(1-r);\nonumber \\
\mbox{for} \; \zeta:&& \vspace{0.4cm} \mu =\frac{1}{2}(1+p); \; \; \tilde \mu =\frac{1}{2}(1-p).
\label{val-pr}
\eeqa

Let us denote by $Q_s$ the following expression
\begin{equation}
Q_s(\rho,\, \zeta):={\rm Tr}(\rho^s\zeta^{1-s}).
\label{qs}
\end{equation}
The functions $Q_s$ are the quantum analogues of the classical R\'enyi overlaps discussed in Ref. \cite{Fuchs} as being distinguishability measures.

One can show that in the case of two qubits, one obtains \cite{Calsa}
\beqa
Q_s(\rho,\, \zeta)&=&\bigg( \lambda ^s\mu^{1-s}+\tilde \lambda ^s\tilde \mu ^{1-s}\bigg) \cos^2\bigg( \frac{\theta}{2}\bigg)\nonumber \\
&&+\bigg( \lambda ^s\tilde \mu^{1-s}+\tilde \lambda ^s\mu ^{1-s}\bigg) \sin^2\bigg( \frac{\theta}{2}\bigg),
\label{qs-explicit}
\eeqa
$\theta $ being the angle between the two Bloch vectors $\vec r$ and $\vec p$. In Ref. \cite{Calsa} the authors have analyzed the quantum Chernoff bound of qubits in the particular case $r=p$ and $\theta =\pi /2$ and shown that in this case $\min_{s\in [0,1]}Q_s$ is obtained for the same value of $s$ regardless of $r$, namely for  $s=1/2$.

\subsection{Quantum Chernoff bound for the output states of the cloning machine. Numerical results}

Further we will investigate the quantum Chernoff bound for the imperfect clones described in Section 2.
Consider that the initial state to be cloned is a mixed qubit state described by the density operator $\rho =1/2\;  (I+\vec r\cdot \vec \sigma )$. Suppose that we have $N$ identical copies of this state. We apply the $N\to M$ optimal universal cloning machine given by the unitary operator of Eq. (\ref{tr-unitara}) and obtain $M$ (with $M\ge N$) output states described by the density operator $\rho_{out}$ of Eq. (\ref{stare-clonata}). An equivalent expression of the output state can be written in terms of the Bloch vector:
\beq
\rho_{out} =\frac{1}{2} \bigg[ I+\frac{N\left( M+2\right) }{M\left( N+2\right)}\; \vec r\cdot \vec \sigma \bigg].
\eeq
We denote $\vec p=\frac{N\left( M+2\right) }{M\left( N+2\right)}\; \vec r$.

Let us investigate the quality of the clone by evaluating how close is the output state to the desired initial one. We propose to employ the quantum Chernoff bound as a measure of distinguishability between these states.
The angle $\theta $ between the Bloch vector of $\rho $ and the one of $\rho_{out}$ is equal to zero. According to Eq. (\ref{qs-explicit}), we obtain the R\'enyi overlaps
\beq
Q_s(\rho,\, \rho_{out})= \lambda ^s\mu^{1-s}+\tilde \lambda ^s\tilde \mu ^{1-s},
\eeq
where the eigenvalues $\lambda $, $\tilde \lambda $, $\mu $, $\tilde \mu $ are given by Eqs. (\ref{val-pr}). The quantum Chernoff bound (\ref{xi}) between $\rho _{out}$ and $\rho $ becomes
\beqa
\xi_{QCB}(\rho,\, \rho_{out})&=& -\ln \; \min_{s\in [0,1]}Q_s \nonumber \\
&=&-\ln \; \min_{s\in [0,1]} \bigg\{ \frac{1}{2}\, (1+r)^s\, \bigg[ 1+\frac{N\left( M+2\right) }{M\left( N+2\right)}\; r\bigg]^{1-s}\nonumber \\
&&+\frac{1}{2}\, (1-r)^s\, \bigg[ 1-\frac{N\left( M+2\right) }{M\left( N+2\right)}\; r\bigg]^{1-s}\bigg\}.
\label{chern-b-clon}
\eeqa

\begin{figure}
\centering
\includegraphics[width=10cm]{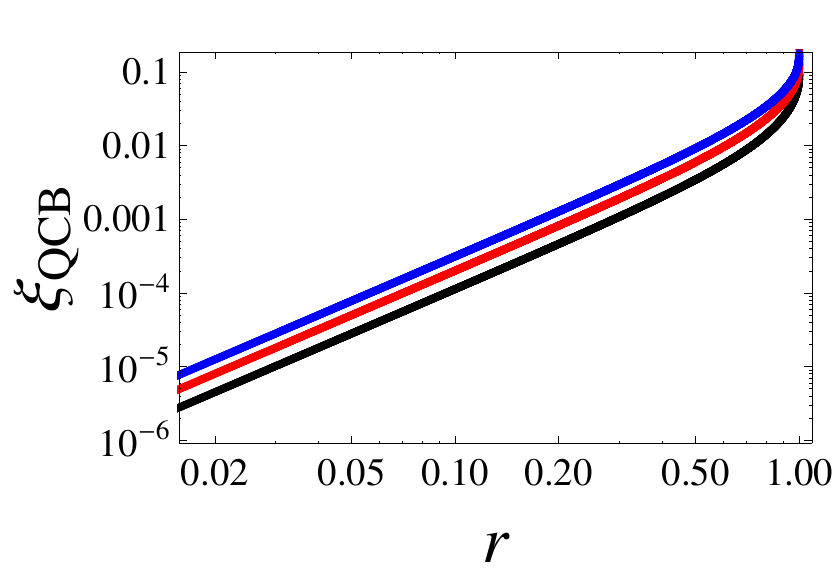}
\vspace{0.5cm}
\caption{The efficiency of quantum cloning given by quantum Chernoff bound in the case of a cloning machine described by $N=2$ and $M=5; 10; 50000$ (from bottom to top) as a function of the length of the Bloch vector of the initial state $r$. }
\label{n-ct}
\end{figure}

\begin{figure}
\centering
\includegraphics[width=10cm]{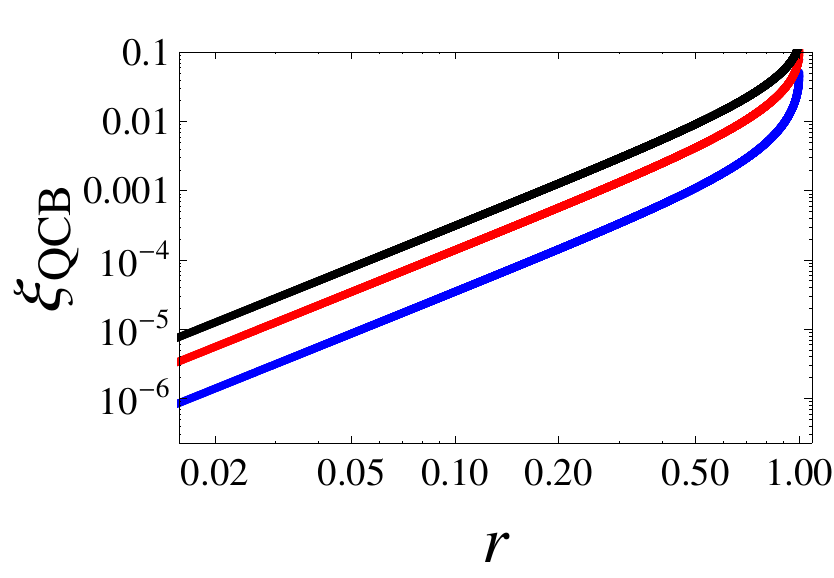}
\vspace{0.5cm}
\caption{The quantum Chernoff bound in the case of $M=10^6$ and $N=2; 4; 10$ (from top to bottom) as a function of the length of the Bloch vector of the initial state $r$. }
\label{m-ct}
\end{figure}

\begin{figure}
\centering
\includegraphics[width=10cm]{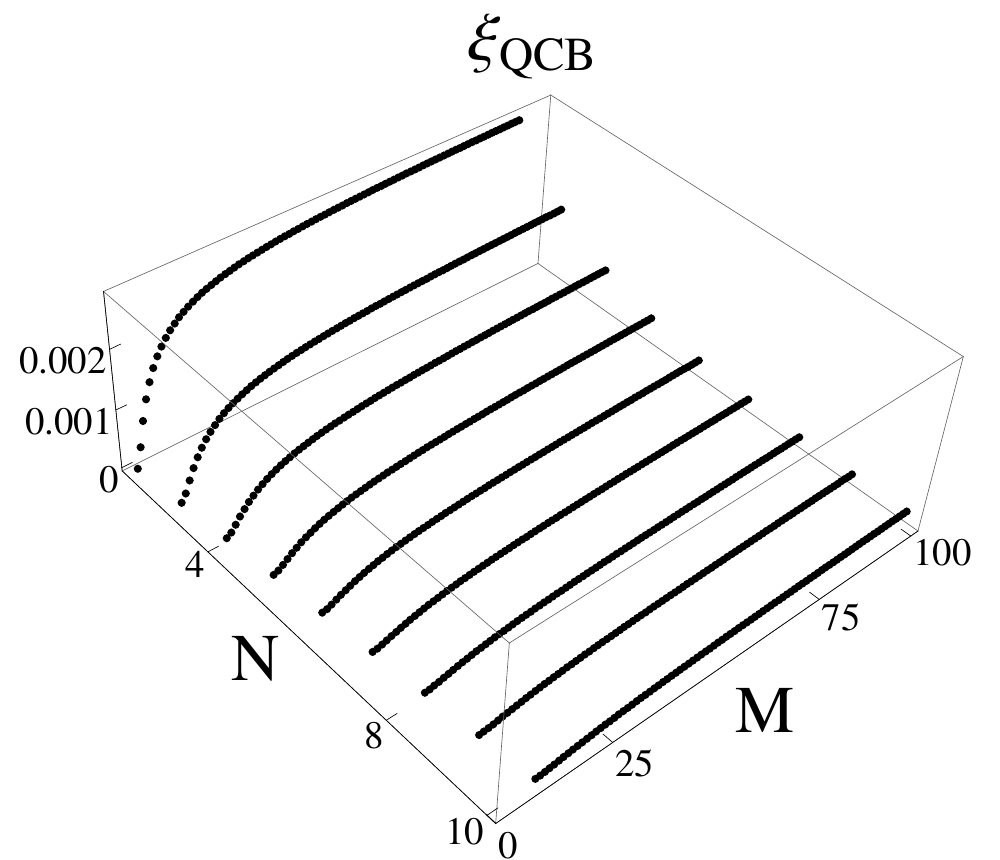}
\vspace{0.5cm}
\caption{The quantum Chernoff bound in terms of $N$ and $M$ for the fixed state characterized by $r=0.3$. We used the condition $M\ge N$.}
\label{r-03}
\end{figure}

\begin{figure}
\centering
\includegraphics[width=10cm]{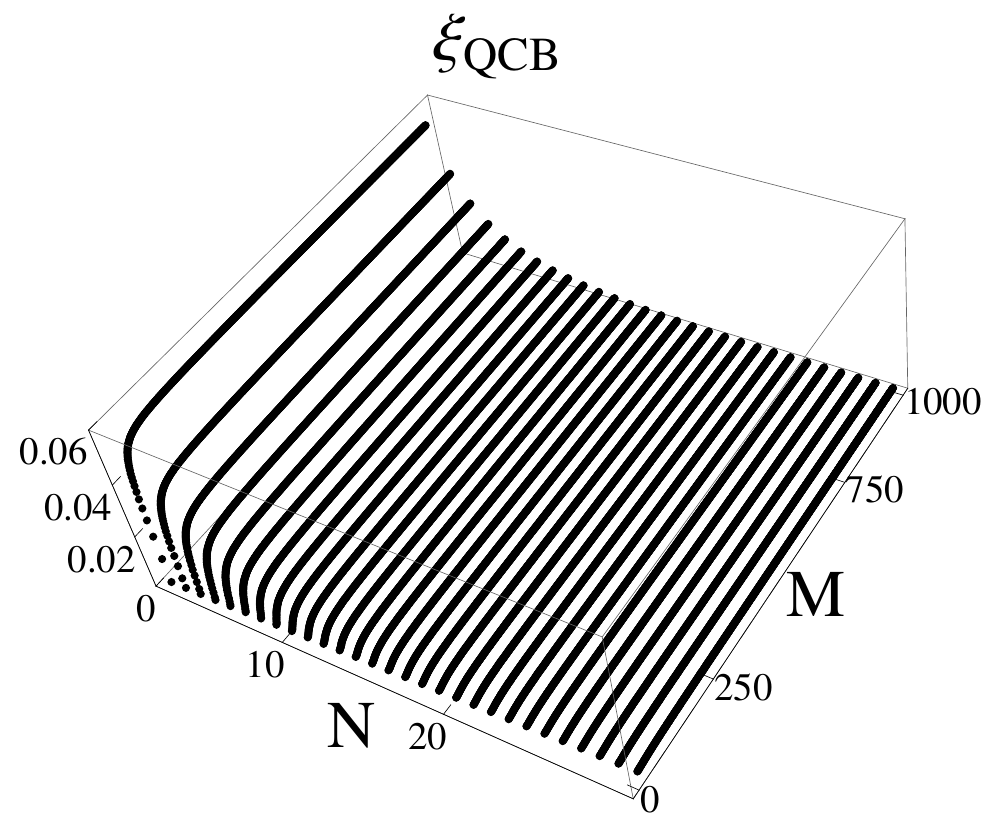}
\vspace{0.5cm}
\caption{The quantum Chernoff bound in terms of $N$ and $M$ for the fixed state characterized by $r=0.9$ (with the condition $M\ge N$). }
\label{r-09}
\end{figure}

In order to find the minimum over $s$ in Eq. (\ref{chern-b-clon}) we need to perform numerical simulations.
One notices from Eq. (\ref{chern-b-clon}) that the quantum Chernoff bound depends on three variables: the number of the identical copies of the state $\rho $ to be cloned $N$, the number of the imperfect clones $\rho _{out}$ generated $M$, and the length of the Bloch vector of the initial state $r$. We perform the numerical simulation for obtaining the quantum Chernoff bound in the following three cases:
\begin{itemize}
\item we keep $N$ = constant. This analysis is illustrated in Fig. \ref{n-ct}: for $N=2$ we plot the quantum Chernoff bound for three values of $M$: 5, 10, and 50000. The quantum Chernoff bound $\xi_{QCB}(\rho,\, \rho_{out})$ increases when $M$ increases.
\item we consider $M$ = constant. The behavior of quantum Chernoff bound for $M=10^6$ is shown in Fig. \ref{m-ct} for three values of $N$, namely 2, 4, and 10. The quantum Chernoff bound decreases as long as $N$ increases.
\item we take a fixed state, i.e. $r$ = constant, and study the behavior of the quantum Chernoff bound in terms of $N$ and $M$. Figures \ref{r-03} and \ref{r-09} show the plots of quantum Chernoff bound in the case when $r=0.3$ and $r=0.9$, respectively.
\end{itemize}

\section{Conclusions}

In this paper we have exploited the quantum Chernoff bound in order to evaluate the efficiency of quantum cloning of mixed qubit states. We have analyzed the imperfect clones generated by the $N\to M$ cloning machine. The quantum Chernoff bound has been used as a measure of distinguishability between the imperfect clones and the initial mixed qubit state.

The quantum Chernoff bound is obtained by computing a minimum over a variable $s$ of the function $Q_s$ defined by Eq. (\ref{qs}). This minimum is determined only by numerical simulations.
We have performed three different kinds of comparisons: firstly we have kept $N$ constant and have shown that the quantum Chernoff bound $\xi_{QCB}(\rho,\, \rho_{out})$ increases with $M$. Secondly, we have considered the number of the output imperfect clones $M$ to be constant. We have noticed that the quantum Chernoff bound decreases as long as $N$ increases. The third case analyzes the behavior of quantum Chernoff bound in terms of both $N$ and $M$ in the case of a given state ($r$ = constant).

\ack
This work was supported by the Romanian National Authority for Scientific Research through Grant PN-II-ID-PCE-2011-3-1012 for the University of Bucharest.

\end{document}